# Occultation Evidence for a Satellite of the Trojan Asteroid (911) Agamemnon

Bradley Timerson[1], John Brooks[2], Steven Conard[3], David W. Dunham[4], David Herald[5], Alin Tolea[6], Franck Marchis[7]

1. International Occultation Timing Association (IOTA), 623 Bell Rd., Newark, NY, USA, btimerson@rochester.rr.com

2. IOTA, Stephens City, VA, USA, jbrooks@johnsweb.com
3. IOTA, Gamber, MD, USA, steve.conard@comcast.net
4. IOTA, KinetX, Inc., and Moscow Institute of Electronics and Mathematics of Higher School of Economics, per. Trekhsvyatitelskiy B., dom 3, 109028, Moscow, Russia, david.dunham@kinetx.com
5. IOTA, Murrumbateman, NSW, Australia, DRHerald@bigpond.net.au
6. IOTA, Forest Glen, MD, USA, alintolea@gmail.com
7. Carl Sagan Center at the SETI Institute, 189 Bernardo Av, Mountain View CA 94043, USA, fmarchis@seti.org

**Corresponding author**
Franck Marchis

Carl Sagan Center at the SETI Institute

189 Bernardo Av

Mountain View CA 94043

USA

fmarchis@seti.org

**Keywords**: Asteroids, Binary Asteroids, Trojan Asteroids, Occultation

**Abstract:**

On 2012 January 19, observers in the northeastern United States of America observed an occultation of 8.0-mag HIP 41337 star by the Jupiter-Trojan (911) Agamemnon, including one video recorded with a 36cm telescope that shows a deep brief secondary occultation that is likely due to a satellite, of about 5 km (most likely 3 to 10 km) across, at 278 km ±5 km (0.0931″) from the asteroid's center as projected in the plane of the sky. A satellite this small and this close to the asteroid could not be resolved in the available VLT adaptive optics observations of Agamemnon recorded in 2003. The outline of Agamemnon is fit well by an ellipse with dimensions 190.6 ±0.9 km by 143.8 ±1.5 km. The angular diameter of HIP 41337 was found to be 0.5 ±0.1 milli-arcsec. After (624) Hektor, this could be the second Jupiter Trojan asteroid known to possess a small satellite.

1. Introduction

As early as 1977 (Dunham and Maley, 1977), claims were made that asteroids probably had satellites, based on observations of occultations of stars by asteroids. But these early claims, mostly based on visual observations with no recordings of the event, were dismissed by most astronomers at the time. One intriguing claim, based on independent observations by two observers at separate locations during a 1980 occultation by the asteroid (216) Kleopatra (Dunham, 1981), was found to be consistent with the currently known orbital elements of one of the now known two satellites of that asteroid (Descamps, et al., 2011), that is, the position of the 1980 secondary event closely matched the projection of the now-known orbit of the satellite in the plane of the sky as computed for the 1980 event. By the end of 2006, 42 suspected satellite occultations had been reported out of a total of over 1029 asteroidal occultations, about 1 for every 25$^{th}$ occultation, but only 9 of these were regarded as reasonably credible at the time (Maley et al., 2007). In 1982, the first video recording was obtained at Meudon Observatory of a "secondary occultation" that was likely caused by a satellite of (146) Lucina (Arlot, et al., 1985). But the field of view for that observation was too narrow to include any other field stars bright enough to be recorded. In this paper, we report a good video observation of a secondary occultation that gives strong evidence that (911) Agamemnon, probably the second-largest Jupiter Trojan asteroid, likely has a small satellite.

The history of asteroidal occultation observations was reviewed in (Timerson, 2009). Successful predictions (Preston, 2013) and observations have increased dramatically, especially since 1997, aided by high-accuracy star catalogs and asteroid ephemerides (Dunham, et al, 2002).

Asteroidal occultations are now usually video recorded using small sensitive security cameras attached to telescopes; the techniques and equipment needed to make these observations are described in the IOTA observer's manual (Nugent, 2012), with basic information also given by (Degenhardt, 2009). Planning software called *OccultWatcher* allows observers to space themselves across the predicted path of the occultation to gather as many unique chords as conditions allow (Pavlov, 2012a). Observations are reported to a regional coordinator who gathers these observations and uses a program called *Occult4* (Herald, 2012) to produce a sky-plane profile of the asteroid at the time of the event (Timerson, 2013). These asteroidal occultation data are officially deposited and archived, and made available to the astronomical community through the NASA Planetary Data System (Dunham, et. al., 2012).

## 2. Occultation Results

### 2.1 Observing stations

On 2012 Jan 19 at 11:31 UT asteroid 911 Agamemnon occulted the V magnitude 8.0 star HIP 41337 = SAO 60804 = BD= +37° 1857 = HD 70920, spectral type K0, in the constellation of Lynx over a path which moved onshore along the mid-Atlantic USA, passed through the central Great Lakes, Canada, and Alaska. The maximum duration was predicted to be 10.4 seconds on the basis of the AKARI AcuA spherical diameter of 185 km (Usui et al, 2011). For this event, 5 observers set up telescopes located at 8 sites across the predicted path of the occultation, using 29.97 frames per second NTSC video to record the event. Five well-spaced chords were obtained across Agamemnon, while two stations reported no occultation, and one station mal-functioned. One observer (S. Conard) used a 36cm Schmidt-Cassegrain telescope at his Willow Oak Observatory in Gamber, Maryland, recording the analog video signal from the video camera directly to a lossless .avi file on a computer, while the other 4 positive observations were made with 0.5 cm binocular-based video systems called "mighty mini" systems (Degenhardt, 2009). The video was recorded digitally to MiniDV tapes using camcorders, and the digital files were later captured in a lossless manner to .avi files via a Firewire interface to a computer using Windows Moviemaker (thus, all of the video recordings of the occultation were not compressed, to avoid the distortions that compression can introduce). Two of these small systems were pre-pointed to the occultation altitude and azimuth, and run as stationary (non-tracking) remote unattended stations, with the video recorders turned on and off at the right time with timers, set up by Dunham. No filters were used for the observations. The cameras used are more sensitive in the red than visual observation, with some sensitivity in the near-infrared range. The occultation timings were determined by analysis of the .avi video files using Limovie (Miyashita, 2008), Tangra (Pavlov, 2012b), and Occular software (George, 2009).

The observations are detailed in Table 1. After the Location column, distances are given from the predicted central line, with positive values northeast of the central line. In the events column, "miss" indicates that no occultation occurred at that station. Most of the video recordings used video time inserters that wrote the UTC time on each video frame; the times for them should be accurate to 1 frame (±0.03s). However, the two successful remote stations, chords 3 and 4, determined UTC from the camcorder clock as calibrated before and after the occultation with short GPS video time-inserted recordings using an "IOTA-VTI" video time inserter; their times should be accurate to 2 frames (±0.06s). The last column gives the signal-to-noise (S/N) ratio for the positive observations. For the miss stations, the S/N is less important, and more difficult to measure, without an occultation period. The station locations, and the predicted path (Preston, 2012) are shown in Figure 1.

**[INSERT Table 1 Figure 1]**

## 2.2 Stellar diameter, Fresnel diffraction

The lightcurves for all observed events were 'gradual', indicative of the effects of stellar diameter and/or Fresnel diffraction. The star is not listed in either the CHARM2 (Richichi et al. 2005) or CADARS (Pasinetti Fracassini et al. 2001) catalogues of stellar diameters. However its apparent diameter can be estimated using either the B or V magnitude with the K magnitude (van Belle, 1999). The V magnitude derived from the Hipparcos Vt and Bt magnitudes is 7.79, while the B magnitude is 9.02. The K magnitude from the 2MASS catalogue is 4.94. The star is not known to be a variable star, nor a giant or supergiant star. Accordingly the stellar diameter estimated using equations (6) and (7) of van Belle is about 0.6mas. Figure 2 shows our analysis of the lightcurves for the occultation by the primary body to determine the stellar diameter. The average of the best fit in each of the 10 lightcurves (optimized by aiming for a minimum in chi-squared) is 0.46 +/- 0.03mas. But the Conard lightcurve has by far the highest S/N; the gradual nature of the occultation events is barely noticeable in the other recordings. Consequently, we prefer to use only Conard's observation, which gives a stellar diameter of 0.5 ±0.1mas – fully consistent with the estimated diameter. We have neglected limb darkening in the analysis; including it could increase the stellar diameter. At the 4.1 AU distance to Agamemnon, 0.5mas subtends 1.5 km, almost 4 times the Fresnel diffraction fringe spacing, which is about 400m at Agamemnon's distance. This fact, and the consistent result that we obtained neglecting diffraction, we believe justifies our neglect of the latter effect; including Fresnel diffraction in the analysis would not significantly change the results, considering the quality of the observations.

[INSERT Figure 2]

## 2.3 Primary Chords and Shape

The resulting chords and least squares ellipse from *Occult4* are shown in Figure 3. The chords are well-fit with an ellipse with dimensions of 190.6 ± 0.9 x 143.8 ± 1.5 km; the RMS of the fit is 1.9 km. This gives an average diameter of 165.6 km, in good agreement with the diameter of 167 ±4 km determined for Agamemnon by the Infrared Astronomy Satellite (IRAS, see Tedesco et al. 2002), but smaller than the 185 ±3 km size given in AKARI AcuA (Usui et al. 2011) that was used for the predicted path shown in Figure 1. This discrepancy could be due to the rotation of the asteroid. No photometry is known of Agamemnon in January 2012, but a month later, the asteroid was observed and its rotational lightcurve was determined (French et al., 2012); the period was found to be $6.59^h$ ±$0.01^h$. The small amplitude in the plot on p. 185 of (French et al., 2012), about 0.02 mag. while it was as large as 0.2 mag. in other years, shows that one of Agamemnon's poles was pointed roughly towards Earth in early 2012, so it is not possible to obtain an estimate of the line-of-sight dimension from the lightcurve. Additionally, no shape model has been determined yet for (911) Agamemnon, and the asteroid's albedo cannot be calculated since its magnitude was not measured at the time of the occultation.

[INSERT Figure 3]

## 2.4 A secondary event

*A very short duration occultation, about 10 seconds after the star reappeared, was noticed in Gamber observations by S. Conard .* The lightcurve showing both occultations is shown in Figure 4. The video was analyzed using two independent tools for measuring analogue video – LiMovie, and Tangra. Figure 5 shows the lightcurve of the secondary occultation in more detail. Close inspection of the lightcurve shows that the star did not completely disappear; about 5% of the star's light remained at the bottom. Also the slope of the disappearance differs from that of the reappearance, consistent with the

satellite not being perfectly spherical. A synthetic light curve was generated using simple modeling on the basis of a uniform 5mas stellar disk. Fig 6 shows the synthetic light curve overlaying the observed light curve, while Fig. 7 shows the model of the satellite used to generate the synthetic light curve. The model indicates the satellite has a diameter of about 5km, but the real size could be anywhere from 3 to 10 km considering the uncertainty on the measurement and the possible elongated shape of the satellite. Again, this analysis neglects Fresnel diffraction. The satellite must have been at least 1.3 km across in order to cover almost all of the star's disk, and that is 3 times the Fresnel diffraction scale, so we believe that neglecting diffraction is justified; this is not like the geometry of small trans-Neptunian objects (TNO's) 25 times farther away, where the diffraction scale is much larger, and necessary to consider (Roques et al, 1987). With only one chord, the real size is unknown. The star could have been occulted by a 6-km mountain on the edge of a much larger object. It could even extend as far south as 36 km before running into Tolea's chord, but a satellite size that large is extremely unlikely, and probably ruled out by the adaptive optics observations described below. This possible satellite is located at a separation of 0.0931 arcsec from the main body at a position angle of 93.8°. At the time, Agamemnon was 4.115 A.U. from Earth, giving a sky plane separation of 278 km.

**[INSERT Here Fig. 4,5,6,7]**

Conard's observation used a 14-inch (36 cm) SCT and GPS time inserted WAT 902H video camera. It occurred in early morning twilight with the sun 10° below the horizon. The target was at an altitude of 24° above the northwestern horizon, in the direction opposite to that of the Sun. Other causes such as birds or planes have been deemed unlikely because such objects should have been visible in the video due to the twilight conditions (they would have to be large enough to cover the 36cm aperture). Other field stars in the recording did not vary or flicker during the occultations; there was no background wavering or other indication of any terrestrial object passing through the field of view. The faintest stars recorded were $13^{th}$ magnitude, 5 magnitudes fainter than the $8^{th}$-magnitude occulted star. The Earth's shadow was at an altitude of about 250 km in the direction of Conard's observation, so any higher artificial satellites would catch sunlight and be visible, or if they were in the shadow, they would move so fast that they would occult the star for less than a video frame (an artificial satellite in a circular orbit at a typical altitude of 400 km moving at 7.8 km/s would have to be over 200m across to occult the star for one video frame, but the star was occulted for several frames, ruling out even the International Space Station). The other video recordings were analyzed for at least a minute before and a minute after the occultation (or closest approach) time to see if they showed any secondary occultations, and none were found. Some artificial satellites were recorded in the wide (about 2° by 3°) fields of view of the small "mighty mini" systems, which recorded stars as faint as 10th magnitude, but no artificial satellites were in the field near the time of the occultation by Agamemnon. There is a possibility that the secondary occultation was caused by a 5km main-belt asteroid, or a background TNO. We have not tried to compute the inherently very small probability of such serendipitous events, a probability smaller than the probability of an occultation by an object within Agamemnon's gravitational sphere of influence, which extends for over 10,000 km from the asteroid.

Two other occultation observations have been reported for Agamemnon. Both involved two positive chords, one on 21 March 2000 (Nugent, 2001) and the other on 6 May 2004 (Blow, 2004). The 2000 observation yielded an ellipse of 168 x 137.5 km, while the 2004 observation showed a 203.2 x 126.6 km asteroid. Neither showed evidence for a satellite. The elliptical fit parameters and the observations can be found in Dunham et al. (2012).

**3. Search for satellites in adaptive optics observations**

The VOBAD database (Marchis et al. 2006a) which gathered most of the publicly available observations of asteroids recorded with adaptive optics (AO) systems on 8-10m class telescopes and the Hubble space telescope contains three epochs of observations of (911) Agamemnon. Table 2 shows the observing circumstances from the JPL-Horizons ephemeris and the characteristics of these observations collected in January 2003 and July 2003. The data were collected using the Very Large Telescope UT4 and its adaptive optics NACO. The S27 camera was used to directly image the asteroid using its visible light as a guiding reference. The data were collected in the near-infrared using the Ks band filter center at 2.18 μm with a width of 0.35 μm for which the AO correction is optimal for faint Trojan asteroid. Four individual frames were recorded per epoch by jittering the target in the field of view. An estimate of the background was derived from these observations and was subtracted from each frame. The frames were also flat-fielded and a bad pixel removal algorithm was applied to them. The final frame shown in Figure 8 is the result of the co-addition of these 4 frames per epoch.

**[INSERT Table 2 and Figure 8]**

Because of the faintness of the asteroid (V~16 mag), the AO corrected images reached an angular resolution of ~120 mas, twice the theoretical resolution of the telescope in this wavelength range, but still significantly better than the seeing resolution (~1 arcsec) at the time of the observation. We can conclude that the primary of Agamemnon is not resolved on this set of observations, since based on the estimated size derived from the occultation and the distance to the observers, its apparent diameter should be less than 50 mas at the time of the AO observations.

A close inspection of the data presented in Figure 8 shows no evidence for a satellite around Agamemnon's primary. However, if we assume that the reported satellite has a faced-on and circular orbit with a radius of 278 km, it should be located between 68 and 75 mas at the time of the AO observations. An analysis of the AO data based on the method described in Marchis et al. (2006b) indicates that the 3-sigma upper limit of detection is typically ~2 mag at this distance from the primary. In other words, a satellite of Agamemnon could be detected at 70 mas from the primary if its diameter is ~60 km at least.

Furthermore, from the profile of the AO observation data, we infer that the smallest moon detectable from our AO data should be at least at an angular distance of 0.6 arcsec and have a diameter larger than 10 km (corresponding to a Δm of 6.1 mag). AO is currently unable to detect a satellite like the one described in this occultation event with (Δm ~7 mag, D ~ 5km, angular separation of 93 mas). A new technique of observation, or development of larger aperture telescope (e.g. TMT, E-ELT) may be capable of validating the detection of this tiny and nearby moon around the Trojan asteroid (911) Agamemnon.

## 4. Conclusion

Results from a high quality video recording during the occultation by (911) Agamemnon of star HIP 41337 (SAO 60804) on 19 January 2012 shows a secondary event consistent with the presence of a satellite. To date, two known Trojan binary systems have been imaged using adaptive optics systems. (617) Patroclus-Menoetius is a similarly-sized binary L5 Trojan asteroid composed of two D~100 km components (Marchis et al. 2006c; Mueller et al. 2010). From the study of the component orbits of (617) Patroclus-Menoetius, Mueller et al. (2010) estimated its bulk density to be $1.1 \pm 0.3$ g/cm$^3$ suggesting a bulk composition dominated by water ice. (624) Hektor is a large D~225 km primary which possesses a 10-15 km satellite (Marchis et al. 2012). Hektor's satellite mutual orbit is currently unknown, so its density remains unconstrained. With a primary of ~150km and a satellite with a diameter estimated to be 5 km, (911) Agamemnon's bulk density is unconstrained, since only one astrometric position of the satellite has been recorded. Unfortunately, further observations using existing adaptive optics technology and Hubble Space Telescope imaging capabilities are impossible since none of these instruments are capable of detecting the satellite.

The binary nature of two L5-Trojan asteroids, (17635) 2006 BC56 & (29314) Eurydamas, was suggested by Mann et al. (2007) based on a photometric survey which revealed a large lightcurve range (~1 mag) and long rotational periods (<2 rotations per day). From this study, our colleagues estimated the fraction of close pair binaries in the Trojan swarms to be between 6 and 10%. The fraction of wide pair binaries with moons, like (624) Hektor and (611) Agamemnon, remains however unknown due to the limited search studies published for this population of small solar system bodies.

Future telescopes with large 30-40m aperture equipped with adaptive optics may be capable in the future to detect such small ($\Delta m \sim 7$) and close (0.09 arcsec) companions of asteroids, revealing the presence of satellites around large Trojan asteroids. For the moment, this detection, and specifically the confirmation of the genuineness of this satellite, relies on the observation of stellar occultation events. The chances of observing this possible satellite during future occultations would be greatly improved with the deployment of more stations.

**Acknowledgements:** FMA work was supported by NASA grant NNX11AD62G. Dunham's work was partially supported by "megagrant" 11.G34.31.0060 from the Russian Ministry of Education and Science. We thank Anthony George, IOTA, for his help in determining the occultation times from the small "mighty mini" system video recordings, and for his independent analysis for both the angular diameter of the star from the occultation data, and his analysis of the secondary occultation lightcurve, which was consistent with our discussions of them above.

**Tables:**

## Table 1. The occultation observations.

| Chord | Observer | Location | Distance from Centerline (km) | Longitude (degrees west) | Latitude (degrees north) | Elevation (m) | Disappearance Time (UT) | Reappearance Time (UT) | Time Standard | Signal to Noise Ratio |
|---|---|---|---|---|---|---|---|---|---|---|
| 1 | B Timerson | Newark, NY | 238.6 | 77.1183 | 43.0067 | 165 | miss | miss | A | NA |
| 2 | D Dunham | Wrangle Hill, DE | 59.4 | 75.6574 | 39.5743 | 13 | 11:31:42.81 | 11:31:47.50 | A | 0.98 |
| 3 | D Dunham | Summit Airport, DE | 51.4 | 75.7136 | 39.5159 | 19 | 11:31:42.06 | 11:31:48.03 | B | 1.27 |
| 4 | D Dunham | Price, MD | 4.7 | 75.9526 | 39.1077 | 17 | 11:31:39.29 | 11:31:48.95 | B | 1.46 |
| 5 | Centerline | | 0 | | | | | | | |
| 6 | S Conard - primary | Gamber, MD | -26.9 | 76.9517 | 39.4692 | 214 | 11:31:40.61 | 11:31:51.13 | A | 9.4 |
| 7 | S Conard - secondary | Gamber, MD | -26.9 | 76.9517 | 39.4692 | 214 | 11:32:01.32 | 11:32:01.50 | A | 9.7 |
| 8 | A Tolea | Forest Glen, MD | -69.2 | 77.0532 | 39.0192 | 110 | 11:31:40.30 | 11:31:48.67 | A | 1.98 |
| 9 | J Brooks | Winchester, VA | -120.8 | 78.2333 | 39.2667 | 213 | miss | miss | A | NA |

Time Standard: A = GPS time inserted video   B = UTC calibrated time video at remote station, see text

**Table 2:** Observing circumstances (r is the distance between the target and the observer, d is the distance between the target and the sun) and characteristics of the AO observations (FWHM is the angular resolution on the observation).

| Date & Time in UT | V mag | r UA | d UA | elongation deg | phase deg | Filter | exposure time (s) | Airmass | Seeing arcsec | FWHM arcsec |
|---|---|---|---|---|---|---|---|---|---|---|
| 2003-01-13T08:11:30 | 16.01 | 5.53063 | 5.6331 | 79.01 | 10.05 | Ks | 240 | 1.25 | 1.03 | 0.115 |
| 2003-07-16T01:42:04 | 15.96 | 5.57243 | 5.4263 | 92.98 | 10.49 | Ks | 160 | 1.44 | 1.27 | 0.127 |
| 2004-07-24T03:25:08 | 15.81 | 5.59397 | 5.1328 | 112.05 | 9.68 | Ks | 240 | 1.53 | 0.74 | 0.130 |

# Figures:

Figure 1. Map showing the observing stations, circles for positive observations, squares for negative (no occultation). The lines show the predicted path, including the central line, the northern and southern limits, and the limits in case of a 1σ shift in the path's location. The actual shadow width was narrower than the predicted width.

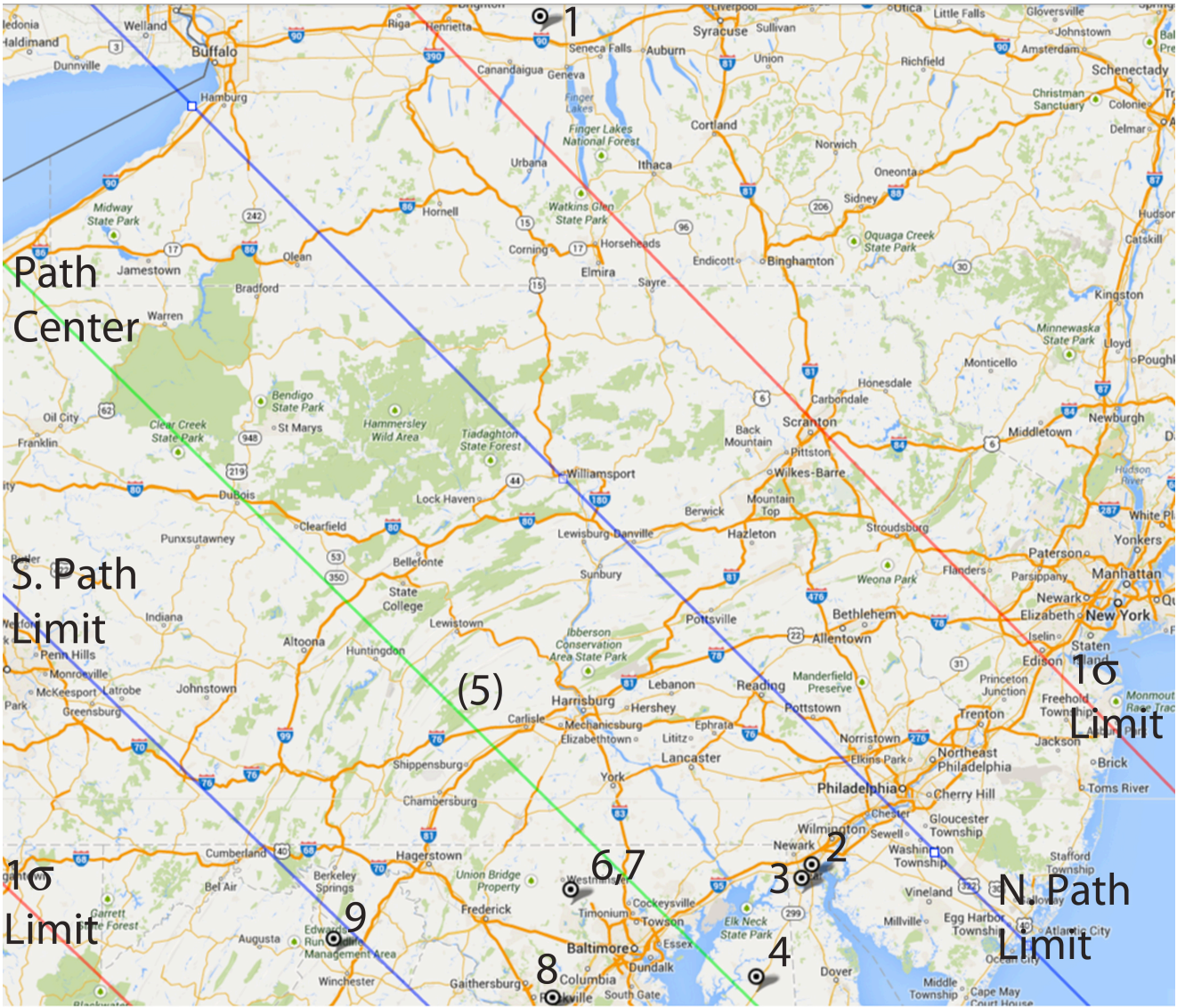

Figure 2. Lightcurves of the occultation events by Agamemnon with the best-fit stellar diameter theoretical lightcurves shown. The observed points at one-frame intervals (each frame is 1/29.97[th] or 0.0333667 second). The abscissa is time increasing from left to right; the tick marks are 0.2 secs apart, and the ordinate is the intensity as measured in a small circle centered on the star in arbitrary units. "D" refers to "disappearance" and "R" to reappearance. The O-C distributions are plotted in the small bar charts's. The noise in the signal has a strong dependence on the signal height. The 'Raw' plot makes no allowance for this dependence, whereas the 'Norm' plot has normalized the residuals having regard to this dependence. The smooth curve is a normal distribution curve to indicate the extent to which the residuals have a normal distribution.

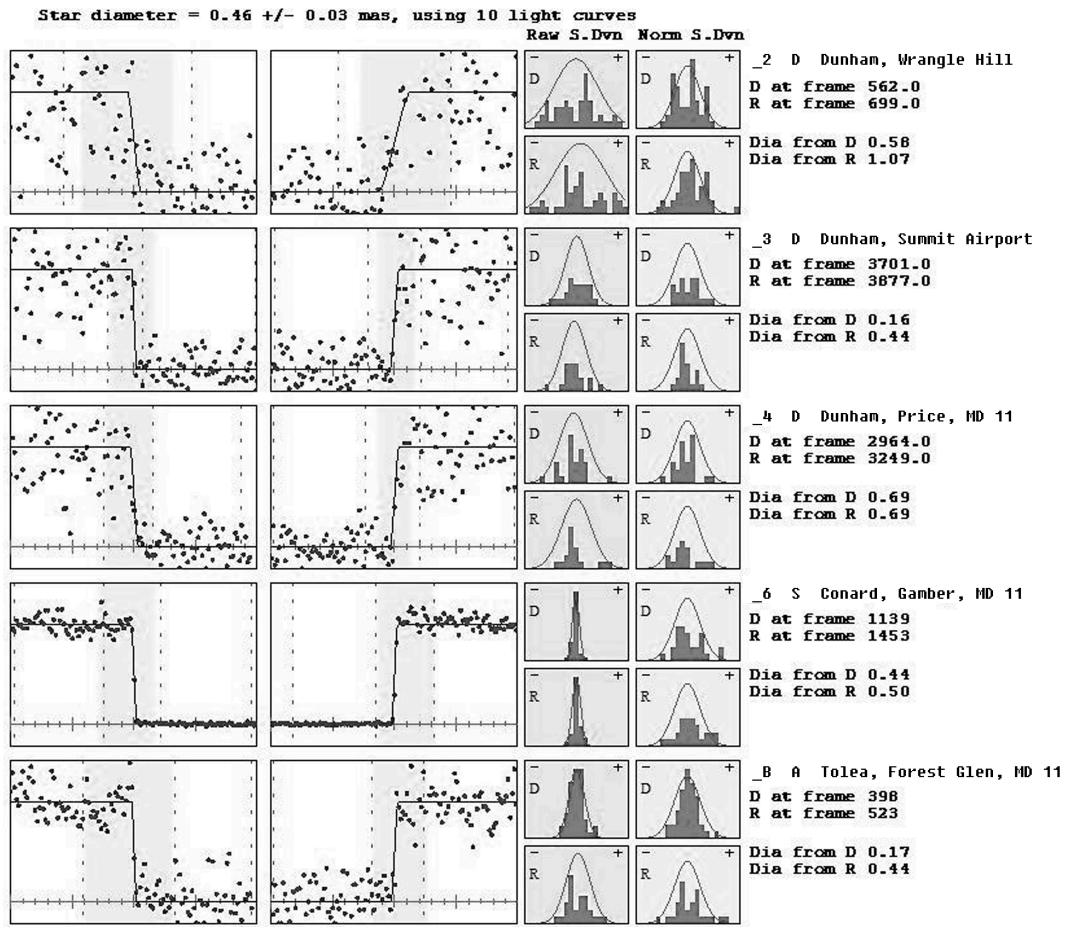

Figure 3. Observed occultation chords projected into the sky plane, including the best least-squares elliptical fit and the possible satellite; no occultation occurred at stations 1 and 9. Disappearances are on the right side of the objects.

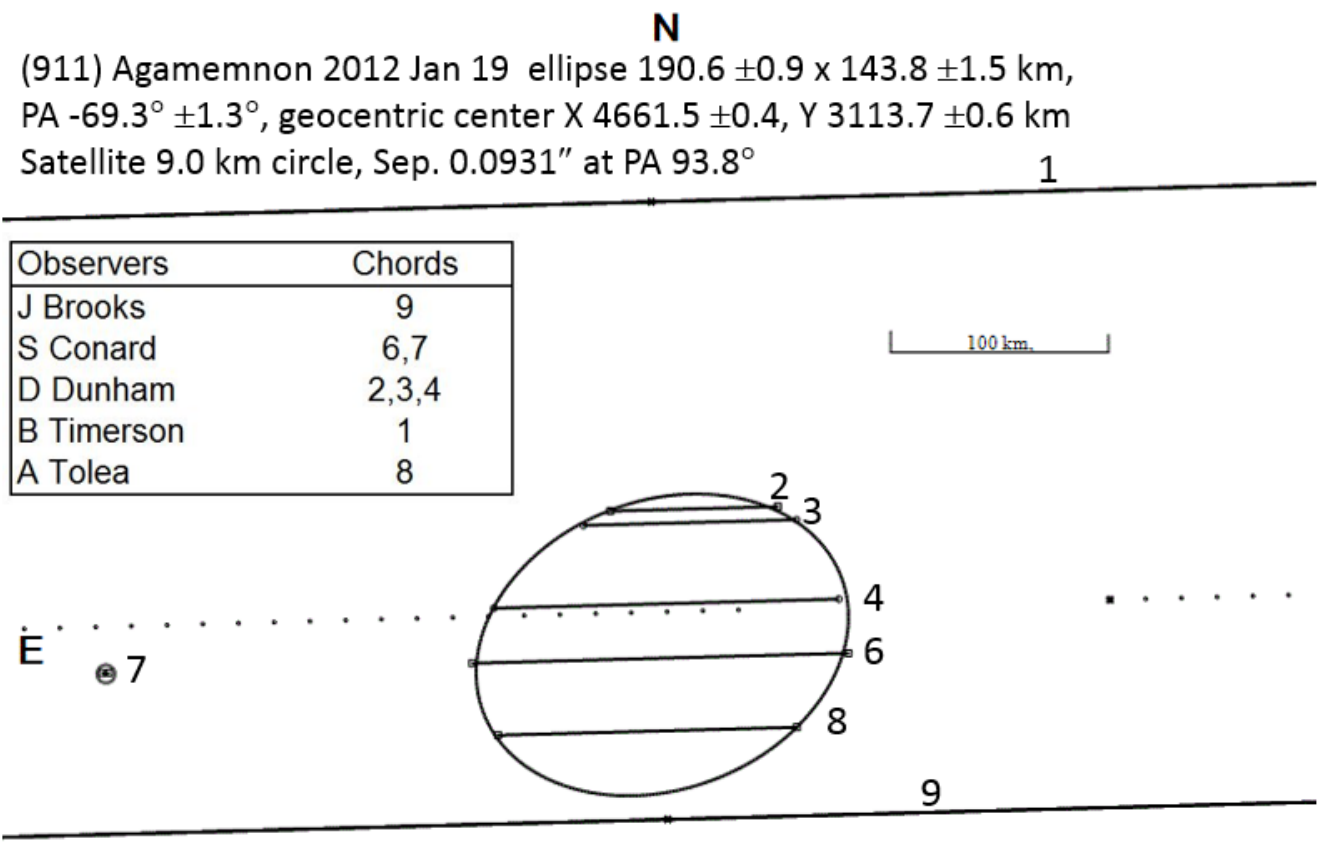

Figure 4. Lightcurve of the occultation recorded at Willow Oak Observatory, showing the secondary occultation near the right side. The abscissa is time measured in video frames (each frame is 1/29.97$^{th}$ or 0.0333667 second) counted from frame 0 = UTC 11h 31m 36.01s and the ordinate is the intensity as measured in a small circle centered on the star in arbitrary units.

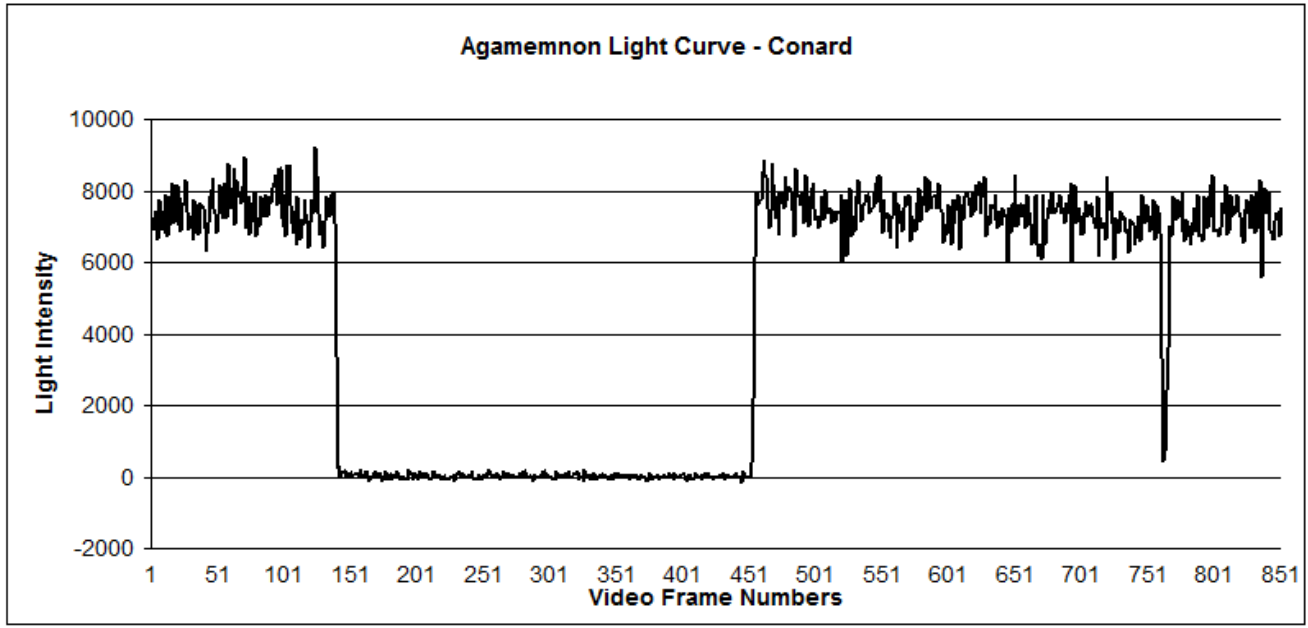

Figure 5. An expanded view of the secondary occultation lightcurve shown in Fig. 4. . The abscissa is time measured in video frames (each frame is $1/29.97^{th}$ or 0.0333667 second) counted from frame 0 = UTC 11h 32m 00.52s and the ordinate is the intensity as measured in a small circle centered on the star in arbitrary units.

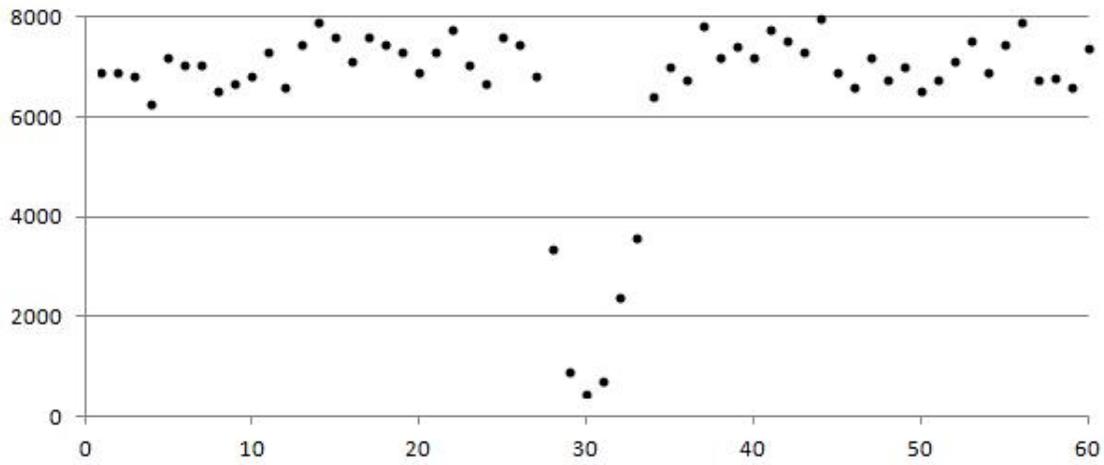

Fig 6 Synthetic light curve overlaying the observed light curve. The synthetic light curve is derived from the model shown in Fig 7.

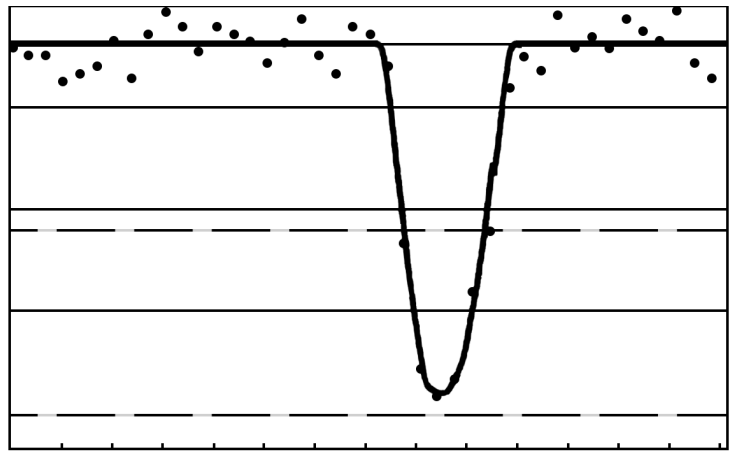

Fig 7. Model of the shape of the occulting object to match the observed light curve. The small disk corresponds to a uniformly illuminated star of 5mas diameter – which is moving from right to left in this diagram. The shape of the satellite cannot be determined by the occultation. The dark region is only a possible fit to the observed light curve The light grey region corresponds to the remainder of the body on the basis of a notional ellipse fitted to the determined region. The scale in km is indicated.

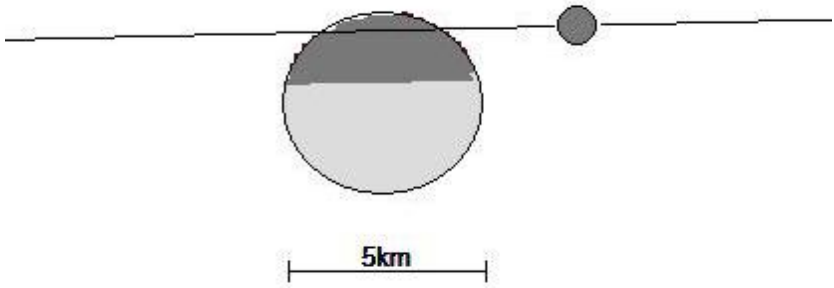

**Figure 8:** AO Observations of (911) Agamemnon with the VLT/NACO instrument in Ks band. No moons are visible from these observations. The Trojan asteroid is not resolved since the angular resolution is 0.12 arcsec, twice the expected angular diameter of the asteroid. The dark circles indicate a face-on and circular orbit with a semi-major axis of 278 km (~0.07 arcsec), corresponding to the projected distance of the satellite detected during this stellar occultation. North is up and east is left on these images.

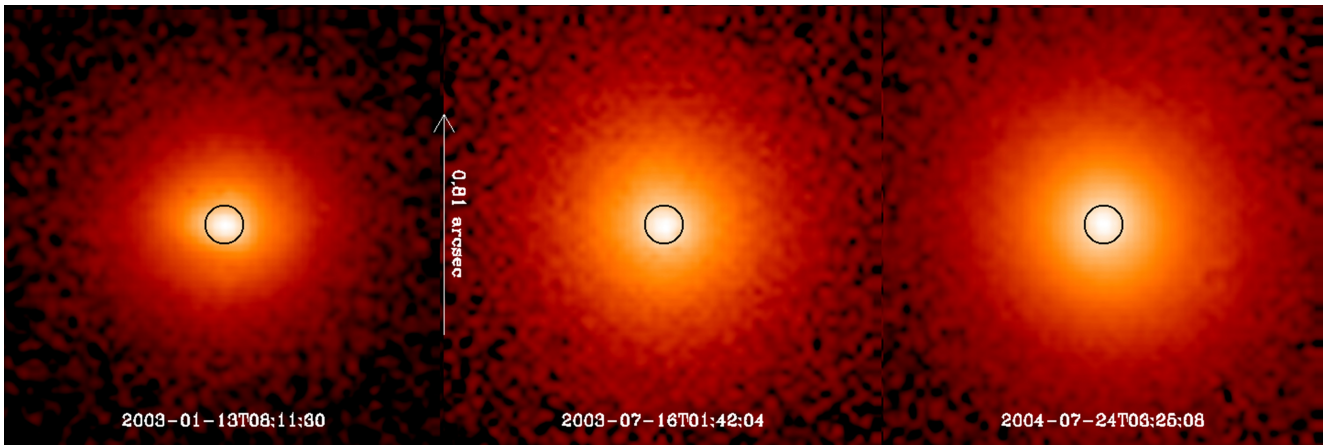